\documentclass[twocolumn]{aastex631}
\usepackage{natbib}

\usepackage{amsthm}
\usepackage{graphicx}
\usepackage[space]{grffile}
\usepackage{latexsym}
\usepackage{amsfonts,amsmath,amssymb}
\usepackage{url}
\usepackage[utf8]{inputenc}
\usepackage{fancyref}
\usepackage{hyperref}
\hypersetup{colorlinks=true,citecolor=blue,urlcolor=blue,pdfborder={0 0 0},}
\usepackage{textcomp}
\usepackage{multirow,booktabs}
\usepackage{caption}
\usepackage{subcaption}
\usepackage{tablefootnote}
\usepackage{xspace}
\newcommand{\R}{\textcolor{black}}
\newcommand{\RR}{\textcolor{black}}
\newcommand{\RRR}{\textcolor{black}}

\date{\today}

\shorttitle{Spectra of a Star's Heterogeneities from its Rotation}
\shortauthors{Berardo et al. 2024}

\begin{document}

\title{Empirically Constraining the Spectra of Stellar Surface Features Using Time-Resolved Spectroscopy}

\author[0000-0001-6298-412X]{David Berardo}
\affiliation{Department of Earth, Atmospheric and Planetary Sciences, Massachusetts Institute of Technology, Cambridge, MA 02139, USA}
\affiliation{Department of Physics and Kavli Institute for Astrophysics and Space Research, Massachusetts Institute of Technology, Cambridge, MA 02139, USA}

\author[0000-0003-2415-2191]{Julien de Wit}
\affiliation{Department of Earth, Atmospheric and Planetary Sciences, Massachusetts Institute of Technology, Cambridge, MA 02139, USA}

\author[0000-0002-3627-1676]{Benjamin V.\ Rackham}
\altaffiliation{51 Pegasi b Fellow}
\affiliation{Department of Earth, Atmospheric and Planetary Sciences, Massachusetts Institute of Technology, Cambridge, MA 02139, USA}

\correspondingauthor{David Berardo}
\email{berardo@mit.edu}

\begin{abstract}
Transmission spectroscopy is currently the technique best suited  to study a wide range of planetary atmospheres, leveraging the filtering of a star's light by a planet's atmosphere rather than its own emission. However, as both a planet and its star contribute to the information encoded in a transmission spectrum, an accurate accounting of the stellar contribution is pivotal to enabling robust atmospheric studies. As current stellar models lack the required fidelity for such accounting, we investigate here the capability of time-resolved spectroscopy to yield high-fidelity, empirical constraints on the emission spectra of stellar surface heterogeneities (i.e., spots and faculae). Using TRAPPIST-1 as a test case, we simulate  time-resolved JWST/NIRISS spectra and \RR{demonstrate that with a blind approach incorporating no physical priors, it is possible to constrain the photospheric spectrum to $\leq$0.5\% and the spectra of stellar heterogeneities to within $\lesssim$ 10\%}, a precision that enables photon-limited (rather than model-limited) science. 
Now confident that time-resolved spectroscopy can propel the field in an era of robust high-precision transmission spectroscopy, we introduce a list of areas for future exploration to harness its full potential, including wavelength dependency of limb darkening and hybrid priors from stellar models as a means to further break the degeneracy between the position, size, and spectra of heterogeneities.


\end{abstract}


\section{Introduction}
\label{sec:intro}

\begin{figure*}[ht]
\begin{center}
    \includegraphics[width=\textwidth]{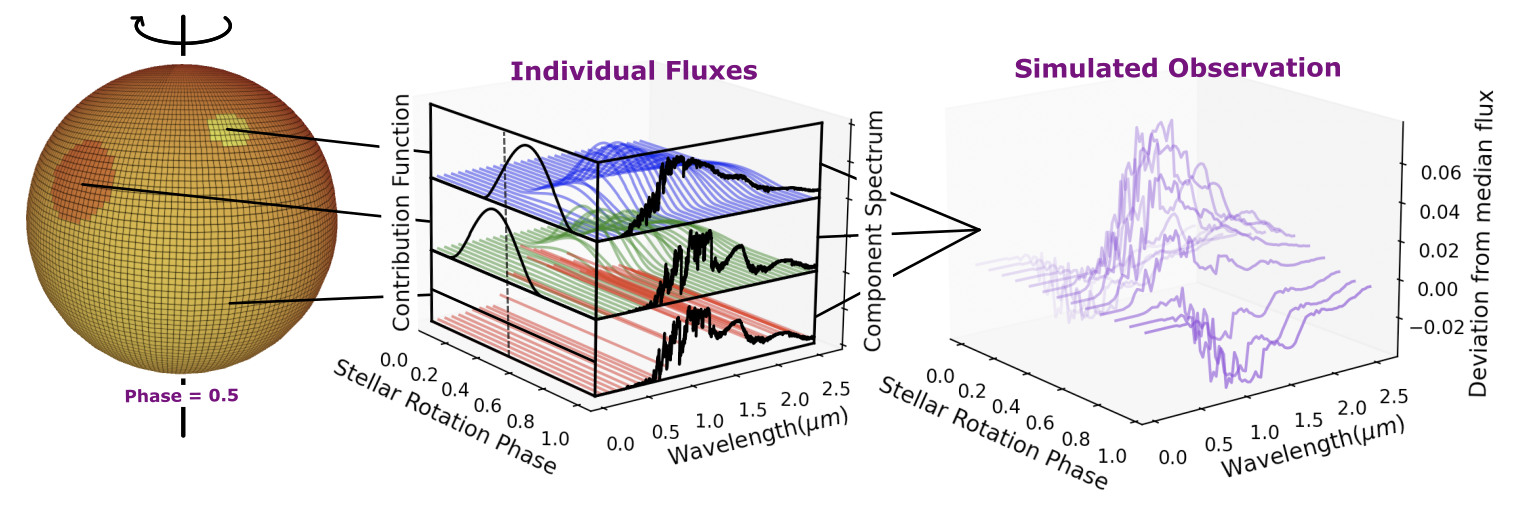}
    \caption{\RR{
    On the information content of time-resolved spectroscopy. 
    (Left) Synthetic stellar surface with two spot features, shown halfway through a stellar rotation (phase = 0.5). 
    (Middle) Individual contribution of each component's time-dependent signal (showing their contribution as the star rotates, as well as effects of limb darkening, size and 2D-projection). 
    (Right) Their summed contributions produce a single time-varying hemisphere-integrated spectrum. We show the deviation from median flux as a function of time, to illustrate the variations due to surface features.}}
    \label{fig:spot_schematic}
\end{center}
\end{figure*}

Transmission spectroscopy was the first technique introduced to study the atmospheres of worlds beyond the solar system \citep{seager:2000,brown:2001}. Today, it is still the most informative \R{technique} for atmospheric studies, as it leverages the light coming from a host star rather than the light directly emitted by the planet itself, which is orders of magnitude fainter. \R{Yet, despite this benefit and subsequent successes achieved, a great deal of work remains in order to make full use of the rich data sets provided by current and upcoming \RR{facilities}, such as JWST \citep{jwst}}.

\R{Currently, the dominant bottlenecks for transmission spectroscopy are associated with imperfections in our opacity models \citep{niraula:2022} and stellar contamination \citep{rackham:2023}.} The current limitations in opacity models result in an accuracy wall preventing constraints on most atmospheric properties beyond $\sim$0.5~dex for all planets but large, hot, and highly-metallic ones \citep{niraula:2023}. Future efforts supporting the standardization of databases, and the improvement of treatments of broadening and far-wings \R{of line profiles}, should mitigate this bottleneck.

Regarding stellar models, \citet{iyer:2020} showed that not accounting for stellar contamination will yield biased inferences of atmospheric properties. \R{More recently, JWST measurements revealed structures in various transmission spectra that predominantly originate from the host stars \citep[e.g.,][]{lim:2023,moran:2023}}. 
Disentangling between the stellar and planetary contribution (i.e., correcting for stellar contamination) is challenging, mostly due to model limitations (i.e., lack of fidelity)\R{,} which can yield a biased contamination correction \citep{sag21}. The lack of fidelity can also result in challenges in inferring the number of components present on the stellar disk \citep{wakeford:2019,garcia:2022}. Fortunately, when stellar models with sufficient fidelity are accessible, the degeneracy between the number of components and their covering fractions can be lifted, leading to an optimal correction of the stellar contamination \citep{rackham:2023}. 

Sufficient fidelity is defined here as follows: with a precision superior or equal to the expected uncertainty associated with the out-of-transit spectra obtained for transit observations in the targeted system. This definition therefore supports returning to a regime of photon-limited studies\R{---}where instruments are used at their maximum potential. \R{The guidance of the report from NASA's Exoplanet Exploration Program Study Analysis Group 21 \citep{sag21} highlights the need for a new generation of stellar models to better understand and model the spectral properties of stellar surface heterogeneities, which in this work we consider to be spots and faculae. Until such models can provide sufficient fidelity,} the only other avenue is to empirically derive the emission spectra of a star's heterogeneities. Doing so would provide the community with a data-driven solution to the stellar-model challenge, as well as benchmarks for the \R{next generation of} stellar models. 

In this letter, we present a framework \R{that leverages time-resolved spectroscopic observations taken across a  stellar rotation cycle to empirically constrain the emission spectra of surface features, such as star spots and faculae, based on their time-varying contributions to the hemisphere-integrated stellar spectrum.}
We focus our injection--retrieval test on M-dwarf stars with properties similar to those of TRAPPIST-1 ($T_\mathrm{eff}$ = 2566\,K\R{; \citealt{agol:2021}}), for which stellar contamination \R{can be more than $10\times$ larger than atmospheric signals} \citep[e.g.,][]{rackham:2018} and the most challenging to correct \R{\citep[e.g.,][]{zhang:2018,wakeford:2019,garcia:2022,lim:2023}}. We present in \autoref{sec:forward_model} the forward model developed to generate the synthetic, multi-wavelength observations of a heterogeneous stellar surface. In \autoref{sec: retrieval framework}, we present the retrieval framework used to assess the extent to which the properties of individual heterogeneities (size, positions, and emission spectra) can be constrained based on a synthetic rotation\R{al} light-curve. In \autoref{sec:injection_retrieval}, we present a series of performance tests of the technique to guide its future usage, as well outline observing strategies amenable to empirical spectral retrieval. Finally in \autoref{sec: discussion}, we consider current caveats and highlight future steps to improve and expand upon this initial framework.

\section{Generating Synthetic Data}
\label{sec:forward_model}

In this section we present the forward model used to generate synthetic time- and wavelength-dependent observations of a heterogeneous stellar surface, which uses  a grid-based stellar surface \R{framework} on which we \R{add} circular heterogeneities (each described by a latitude, longitude,  radius, and temperature) representing circular spots and faculae \citep{montalto:2014}. This is forward model is built upon the model used in \citet{gunther:2022} to simulate the interactions of a heterogeneous star with a debris disk. A graphical depiction is show\R{n} in \autoref{fig:spot_schematic}.

\subsection{Spectral Model}
\RR{We simulate observations for the NIRISS (Near-Infrared Imager and Slitless Spectrograph) instrument \citep{niriss_soss} on JWST \citep{jwst}, which provides an ideal compromise between resolving power \RR{($R{=}700$)} and spectral coverage \RR{(1--2.5\,$\mu m$)}, considering the spectral energy distribution (SED) of stars.} We use the PHOENIX stellar spectral model grid\footnote{\url{https://phoenix.astro.physik.uni-goettingen.de/}} to simulate the emission of an individual surface feature considering their respective viewing angle $\mu$ \citep{husser:2013}. These grids provide adequate \R{wavelength, temperature, surface gravity, and metallicity coverage}  to describe the photospheric background of an M dwarf, as well as surface heterogeneities. For this synthetic forward model, we include a quadratic limb-darkening profile $I(\mu)/I_0 = 1 - c_1(1-\mu) - c_2(1-\mu)^2$ \citep{kipping:2013},  with $c_1 = 0.26$ and $c_2 = 0.4$\R{,} which are obtained using the ExoCTK Limb Darkening Calculator\footnote{\url{https://exoctk.stsci.edu/limb_darkening}}, using the parameters of TRAPPIST-1 in the NIRISS bandpass.

For the photosphere we use a spectral model with a temperature of 2500\,K, a $\log g$ \R{surface gravity} of 5.0, and an [Fe/H] metallicity of 0 (similar to TRAPPIST-1, which has a surface temperature of $2566 \pm26$\,K, a $\log g$ of $5.2396 \pm 0.006$ \citep{agol:2021} and an [Fe/H] metallicity of $0.05 \pm 0.08$ \citep{ducrot:2020}). For heterogeneities, we use spectra corresponding to 2300\,K and 2700\,K (varying $\pm 200$\,K relative to the photosphere, based on Figure 1 of \cite{herbst:2021}) \R{as well as non-physical emission spectra to highlight that our framework can retrieve emission spectra without any a-priori constraints (\autoref{fig:spectra_fits}).} Based on previous NIRISS observations of the TRAPPIST-1 system, we generate Gaussian noise with a scatter of 1.8\% \R{per pixel}, \RR{based on simulated NIRISS observations made using the Pandexo tool \cite{pandexo:2017} and observations presented in \citep{lim:2023}, for an integration time of two minutes.}

\subsection{Spatial Model}

\R{The stellar surface is treated as a rectangular grid running along the longitude and latitude of the surface, having twice as many grid elements in the latitudinal direction. The number of cells (i.e., the resolution) is a free parameter in the model. With these parameters, the projected surface area differs from the true value of $\pi$ by only 0.02\% (i.e., 20\,ppm), well below the injected uncertainty. In this work we consider circular spots and faculae \citep{montalto:2014} as our test models.}

\section{Retrieval Framework}
\label{sec: retrieval framework}

The goal of this work is to demonstrate the ability to characterize arbitrary heterogeneities of a stellar surface and their contribution to the overall stellar spectrum without relying on physical spectral models, which currently cannot provide a sufficient level of accuracy and \RR{thus leads to biased inferences and corrections.} The contribution function of a heterogeneity is non-linear, due to both projection onto the observing plane as well limb-darkening effects, and thus when retrieving these geometric properties of a surface feature we employ standard Markov chain Monte Carlo (MCMC) methods in order to sample the full range of parameter space. 

The observed flux at a given time and wavelength can be expressed as a linear combination of the stellar photosphere and the heterogeneity spectra (scaled by their relative projected surface area):
\begin{equation}
    \label{eq: spectrum breakdown}
    \resizebox{.9\hsize}{!}{$
    \mathrm{Flux}(\lambda,t) = \Lambda_{\mathrm{phot}}(\lambda) + \sum_{i}\left[\Lambda_i(\lambda)-\Lambda_{\mathrm{phot}}(\lambda)\right] \times S_{i}(t)
    $}
\end{equation}
where $\Lambda_{\mathrm{phot}}(\lambda)$ is the (constant in time) spectral signal of the photosphere (\R{including limb-darkening effects}), $\Lambda_i(\lambda)$ is the spectrum of the $i^{th}$ heterogeneity, and $S_{i}(t)$ is the time-varying geometric projection of a heterogeneity, which is a function of its size and position on the stellar surface, as well as any limb-darkening effects. The sum runs over the number of individual heterogeneity features. A graphical depiction of this decomposition is show\RR{n} in \autoref{fig:spot_schematic}.

\RR{With the resolving power now accessible, the inverse problem yielding the properties of stellar heterogeneities is well posed. Each stellar surface feature will have 3 spatial properties (longitude, latitude, size) and $n_\Lambda$ spectral bins for its emission spectra. For $N$ distinct features, the total variables associated with this inverse problem is thus ($3+n_\Lambda)\times N$. In comparison, the data sets being considered have $n_T \times n_\Lambda$ data points, where $n_T$ is the number of exposures. NIRISS observations allow for $n_\Lambda = O(1,000)$ spectral bins, thus as $n_T >> N$ in order to build up the signal-to-noise ratio and sample different phases, the problem entails at least two orders of magnitude more data points than unknown variables, i.e., the problem is well-posed.
}

Part of the problem is linear (for a given set of positions) and thus singular value decomposition (SVD) can be used, leveraging rapid and robust libraries available in Python to estimate the spectral signal ($\Lambda_i$) of each feature in just a few milliseconds. Here, we thus separate the problem into a non-linear MCMC retrieval (the geometric properties of the heterogeneity) and linear retrieval (the spectral signal of the photosphere and individual heterogeneities). In practice, the geometric properties of the surface features are primarily recorded in the white light curve \citep[see ``contribution function'' in Figure 2, and][]{walkowicz:2013,luo:2019}. Thus, light-curve inversion techniques can provide adequate first guesses for the positions, sizes, and numbers of heterogeneities.

\section{Performance Tests}
\label{sec:injection_retrieval}

Given the forward model used to simulate observations described in \autoref{sec:forward_model}, and the retrieval \R{framework} described in \autoref{sec: retrieval framework}, we now describe a series of injection--retrieval tests we use to assess the ability of the retrieval framework to recover stellar surface heterogeneities \RRR{when observing an entire rotation of a test star with a three day rotation period using two-minute sampling.}


\subsection{Fitting for Spectral Components}

In order to test the effectiveness of the \R{fitting framework} in retrieving spectral features of a star, we first perform a series of injection--retrieval tests in an idealized scenario in which we first assume to know the number of heterogeneities, as well as their positions and sizes, attempting to retrieve only the spectral features of heterogeneities and photosphere (the linear part of the retrieval), which represents a best-case scenario and effectively acts as an upper limit on the capability of the framework. 
This can similarly represent a scenario where strong priors have been obtained for the spectral components, based on an analysis of a white lightcurve or a pre-fitting routine which places constraints on the possible heterogeneity configurations.

\begin{figure*}[ht!]
\centering	
\includegraphics[width=\textwidth]{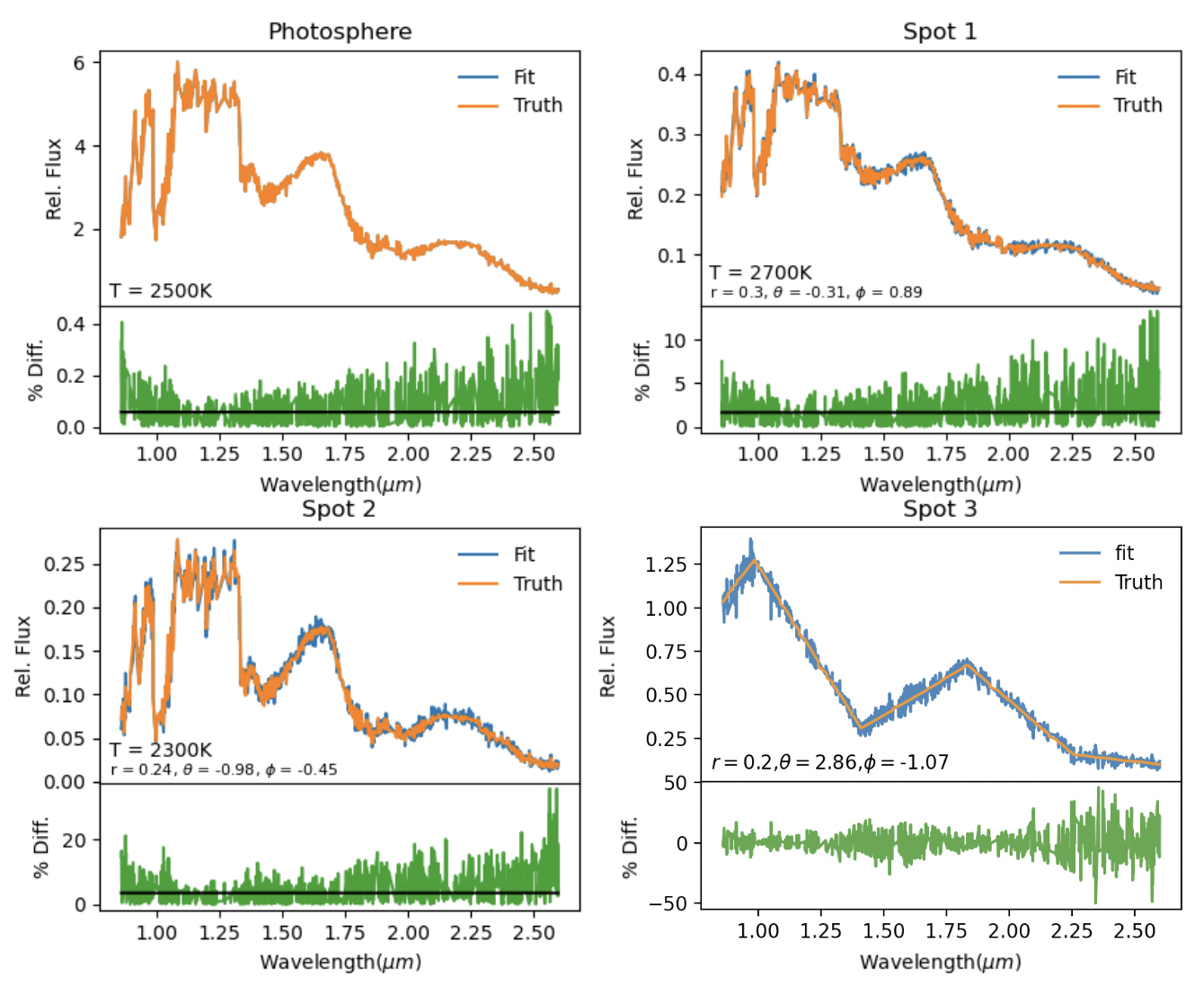}
	\caption{\R{Proof-of-concept for stellar spectra retrieved using time-resolved spectroscopy. Examples of retrieved spectral signals for a photosphere and three spot features on a synthetic star. Their parameters---temperature, size, latitude ($\theta$), and longitude ($\phi$) in radians---are given in the bottom left corner of each panel \RR{and are kept fixed in this first retrieval}. Synthetic and retrieved spectra are in orange and blue, respectively, and residuals at the bottom. For this example, we purposely use a non-physical emission spectrum for spot \#3 to emphasize that the framework reliably retrieve\RR{s} the emission spectra empirically. }}
	\label{fig:spectra_fits}
\end{figure*}

We tested the \R{framework} on a suite of stellar surfaces which are described in \autoref{tab: model features}. The results of these tests reveal that the \R{framework} is able to recover the spectra of heterogeneities to sufficient precisions (i.e., better than the \RR{OOT (out-of-transit) spectrum---see \autoref{sec:forward_model}). For example, the precision achieved on the photospheric spectrum \RR{in this idealized scenario of knowing the spot positions}  is $\le$ 0.1\%. We considered observing conditions similar to those of \cite{lim:2023} for TRAPPIST-1, and estimated OOT uncertainties of $\sim$0.5\% using the Pandexo tool \citep{pandexo:2017}}
---typically based on a ${\sim}2$\,hr integration. The spectra of heterogeneities are \RR{on average constrained to $\lesssim$ 10\%, depending on their sizes and latitudinal position, and more significant deviations are typically due to difficulties in localizing spots, which can be improved by including additional physical information, as discussed in \autoref{sec: discussion}}. We show the results of an example fit in \autoref{fig:spectra_fits}, comparing the individual retrieved component spectra to the spectra used to generate the synthetic observations.

\R{In addition to spots with physically motivated spectra (spots 1 \& 2 in \autoref{fig:spectra_fits}), we also included in that model a completely synthetic and non-physical spectrum (spot 3). We do this to illustrate that our model is able to recover any inherent signal present in the data, without biases from what is assumed to be physical or non-physical. This allows flexibility with respect to inaccuracies that may be present in spectral models used to fit observations.}

The spectra of heterogeneities are less constrained due to their smaller covering fraction\R{s and thus fewer photons.} \R{Their small covering fractions also mean that while the uncertainties associated with their spectra are larger, they contribute less to the total uncertainty budget for the stellar model than the photosphere}. For this reason, we will assess \R{retrieval} fidelity based on the ratio of the uncertainty associated with the retrieved photospheric spectrum and the one associated with the out-of-transit spectrum. 




\subsection{Varying Observation Baseline}

In the previous sections, \R{the} retrieval was performed using simulated observations covering an entire rotation period of the host star. However, in most cases a strong argument must be made to justify the use of high-demand facilities to continuously stare at a single target. In this section, we investigate the ability of the framework to accurately measure the photospheric spectrum of a star when observing \RR{a continuous snapshot of the full rotation lightcurve for less than one full orbit}. Given \R{that a surface feature is only visible for half of the stellar rotation period},  
there exists a strong correlation between the duration of an observation, the \R{phase of the stellar rotation being observed}, and the retrieved uncertainty on the stellar photosphere.

\begin{figure}[hbt!]
	\centering
        \includegraphics[width=\columnwidth]{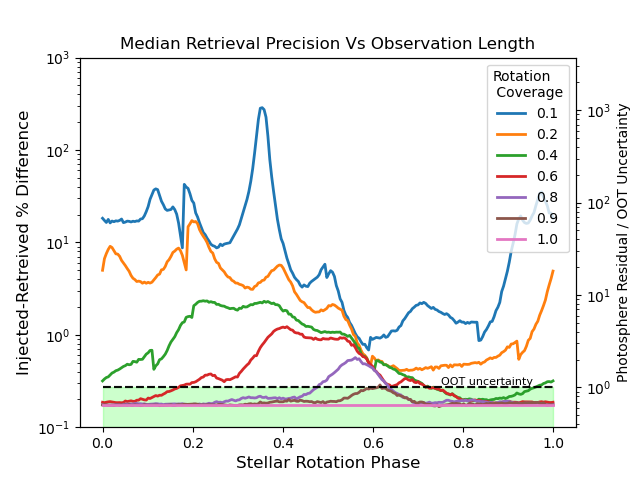}
	\caption{\RRR{Minimum phase coverage of observations to leverage time-resolved spectroscopy. The Y-axis represents the median error on the photosphere during a continuous observation of a star with three spot features. The x-axis shows the starting phase of an observation, while different colors correspond to different durations of a continuous observation (relative to the full rotation period). I.e. The orange line illustrates the median photospheric error when observing for 20\% of a stellar rotation cycle, for an observation beginning at all possible times during a rotation period. The dashed line shows the targeted precision, chosen as the out-of-transit spectrum of a previous NIRISS observation of the TRAPPIST-1 system (``OOT uncertainty'').}}
	\label{fig: shifting_obs}
\end{figure}

\begin{figure*}[hbt!]
	\centering
        \includegraphics[width=7in,angle=0]{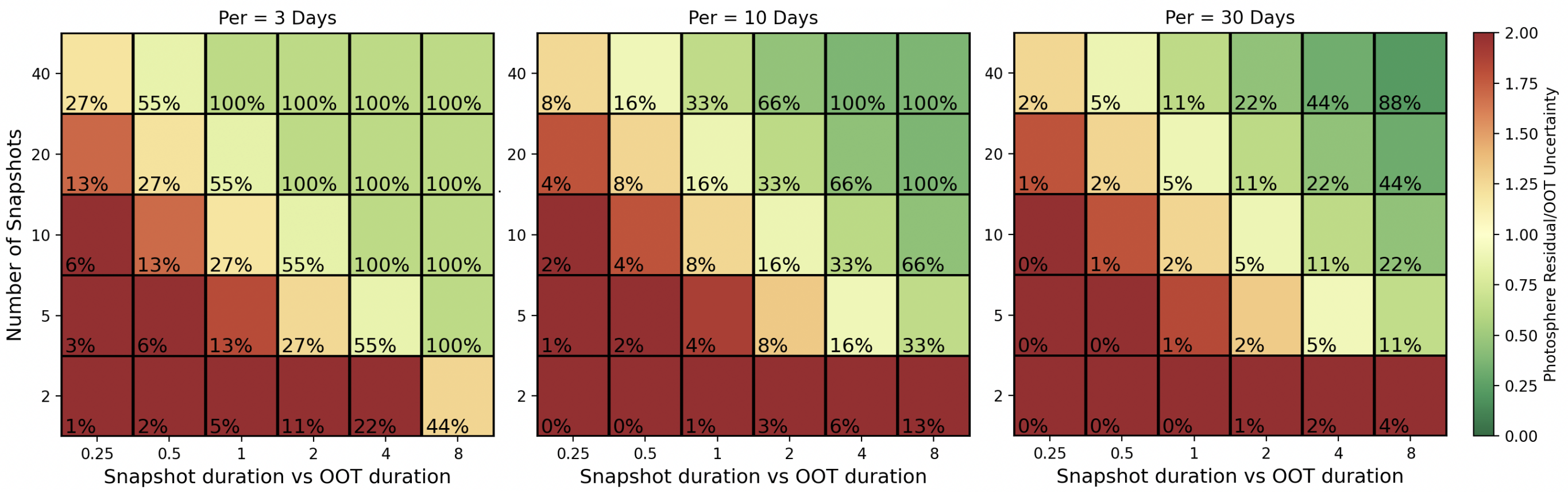}
	\caption{\R{Long-stare vs series of snapshots\RR{, highlighting the effectiveness} of different observing strategies for stars with rotation periods of 3, 10, and 30 days.} \RR{Heat maps show} the median residual on the photosphere spectrum when observing in snapshots (relative to the OOT uncertainty) for three different rotation periods. The percentage in each cell represents the total observation time relative to the rotation period of the star. \R{The snapshot duration is relative to a 2-hr duration that would be typically observed OOT for a transit of planets around TRAPPIST-1.}}
	\label{fig: snapshot_heatmap}
\end{figure*}

We run a similar retrieval as in the previous \RR{section, using the same time sampling of two minutes}, having also chosen: (1) a \R{phase offset} for the longitudinal rotation of the star relative to the observer, and (2) the length of a continuous observation or `stare', defined as a fraction from 0--1 of the stellar rotation period. Selecting a value of one for the viewing fraction represents the analysis done in the previous section, for which the entire stellar rotation was supplied to the fitting routine. These two values define a time series, for which we generate the geometric flux signals attributed to each heterogeneity on the stellar surface and then retrieve their component spectra, the results of which are shown in \autoref{fig: shifting_obs}.

The various curves represent different observation durations. For a given observation duration, the residual signal varies strongly as a function of stellar rotation phase. This is more pronounced for the shorter durations. For example, the residual for an observation covering 0.1 of the stellar rotation can vary from approximately 1\% to over 100\%. For this reason, we find that only a \R{phase coverage} of $\geq90\%$ can reliably constrain the stellar spectra to within the OOT uncertainty (0.5\%). Indeed, while the targeted precision of 0.5\% may be achieved for some configurations with only a 40\% phase coverage, it is not achieved for all (average precision $\sim$1\%).  



\subsection{Series of Snapshots for Slow Rotators}

Covering 90\% of a stellar rotation of TRAPPIST-1 would correspond to a $\sim$72-hr stare at the system \R{based on \RR{its} ${\sim} 3.3$-day period \citep{luger:2017,vida:2017}}, which is both feasible and reasonable for such a high-priority target. Doing so for slow rotators that may have periods up to 30 times that of TRAPPIST-1, however, would be impractical. For such hosts, we show that a series of small stares (``snapshots'') could be used instead (see \autoref{fig: snapshot_heatmap}). In order to reach the targeted precision, we find that snapshots need a minimum duration equal \R{to} the intended OOT integration and sufficient occurrences to sample time-varying contributions of the heterogeneities.

\RR{As the heatmaps in  \autoref{fig: snapshot_heatmap} show, the duration and number of snapshots required to achieve a given SNR are related, offering multiple observational options. As in previous sections, the retrieval here is done assuming to know the true spot positions, which will not be the case in general. However, our intention is to highlight the relative effectiveness of different observing strategies. For example, we find that for a 30-day rotation period a similar precision is achieved for, e.g., 40 2-hr snapshots, 20 4-hr snapshots, 10 8-hr snapshots, or 5 16-hr snapshots. These options correspond to a 10$\times$ lower observation requirement than for a continuous stare, which provides flexibility when selecting observing strategies, taking into account practical considerations such as overheads and slew times.}

\subsection{Reliably Retrieving Heterogeneities' Geometric Properties}

In order to fully test the ability of the \R{retrieval framework} to characterise a heterogeneous stellar surface, we also run a set of retrievals where we use MCMC to estimate the geometric properties (i.e., size, position, limb darkening) of each \R{in}homogeneity in addition to using SVD to estimate their spectral signature. 



\RR{We ran fits for the same set of models described in the previous section, varying the size, position, temperature, and number of spot features present on the stellar surface. We show an example retrieval of a one spot model in \autoref{fig: real fit}, which highlights the strengths and weaknesses of the framework. The framework is able to reliably constrain the longitude of features to within $\pm 5 \degr$. This can be improved by pre-fitting a white-lightcurve, where we sum across all wavelengths. While this discards spectral information, we can then use this constrained longitude as a prior when retrieving the spectral signal of the features. The latitude, in comparison, has a much wider range on average, with an typical uncertainty of $\pm 20 \degr$. This amounts to differentiating only between equatorial or polar spots \citep{luger:2021}.} 

\RR{Additionally, as shown in \autoref{fig: real fit}, the size of the spot feature can similarly vary by up to $50\%$ of its true value. The result of this is a degeneracy between size and amplitude of the spectral contribution of the heterogeneity. However, as also seen in \autoref{fig: real fit}, despite these pitfalls the framework is still able to accurately estimate the spectrum of the stellar photosphere. In \autoref{sec: discussion}, we outline how additional prior information can further constrain the size of a feature, based on global physical constraints on the overall scaling of its spectrum (leveraging the trade-off between feature size and spectral amplitude) to better constrain the physical features of the spot. }




Heterogeneities were allowed to occur anywhere on the stellar surface\R{,} leading to degeneracies where two heterogeneities would overlap and contribute to the overall spectrum jointly. Additionally, we found that without additional information, the latitudinal position of a heterogeneity was difficult to constrain. These issues highlight clear areas for improvement for future work, which we discuss further in \autoref{sec: discussion}. 
\RR{In this unconstrained scenario, the framework often has difficulty in determining precise geometric features of the heterogeneities (position and size), often due to degeneracies between the position and number of features. However despite this, it is still able in most cases to recover the photospheric signal to within 1\%.}







\section{Discussion \& Future Steps}
\label{sec: discussion}

\R{This work presents a potential avenue--and its performance assessment--towards solving the challenges associated with stellar contamination of transmission spectroscopy, in particular addressing one of the main bottlenecks regarding the lack of fidelity of current stellar models. This is in parallel to efforts which aim to identify and correct for stellar activity effects through ground-based high-resolution spectroscopy \citep{mallonn:2018,rosich:2020,perger:2023}, which still requires stellar spectrum templates. Still such initiatives highlight both the communal interest and effort in tackling these issues.}

Leveraging time-resolved spectroscopy to generate empirical emission spectra of stellar heterogeneities will enable a new era of high-precision transmission spectroscopy, in-depth atmospheric studies, and critical benchmarks for the next generation of stellar models.  The upcoming library of empirical emission spectra for stellar surface heterogeneities will be similar in scope to the work of \cite{kesseli:2017} that compiled a library of empirical spectra for various stellar types, with the important distinction that the spectra being considered here are not for disk-integrated features, but rather for `pure' basis components which may be combined with rotational geometry in order to produce accurate spectra for stars with arbitrarily complex surface features. Such a library will not only enable the robust correction of the transit light source (TLS) effect based on out-of-transit measurements \citep{rackham:2018,rackham:2019}, it will also provide important benchmarks for the next-generation of theoretical stellar models \citep[e.g.,][]{Witzke2021,sag21}, and further inform key relationships between the properties of stars and those of heterogeneities, such as between heterogeneity temperatures and sizes, photospheric temperatures, and atomic line-depth ratios. 

\RR{We have demonstrated that under ideal conditions of knowing the spot geometry,} we are able to constrain photospheric spectra at the level of the $\lesssim 1\%$ for the spectra of heterogeneities while
spectra with precisions of $\sim 1\%$ ($\mathrm{S/N} \sim 100$) are used commonly to constrain the fundamental physical parameters of exoplanet host stars \citep[e.g.,][]{wells:2021, delrez:2022, barkaoui:2023, ghachoui:2023, pozuelos:2023, dransfield:2024}. \RR{We have also taken first steps towards are a fully blind retrieval, and have similarly found situations in which the stellar photosphere can be constrained at a similar precision.} 
In terms of absolute flux calibrations, for example, the goal for the X-SHOOTER instrument is $\leq 10\%$ \citep{schonebeck:2014}, while the eventual goal of the JWST calibration program is $1\%$ accuracy for each observing mode \citep{gordon:2022}.
Thus, constraints on component spectra from this technique are on par with current precisions available for integrated disk spectra and will be limited ultimately by the overall precision and accuracy limitations of JWST observations themselves\R{,} providing valuable data-driven benchmarks to inform the next generation of models.

\R{While we have demonstrated the prospects using time-resolved spectroscopy to empirically measure the spectrum of stellar surface features, we highlight that there remain challenges and opportunities to leverage its full potential. We conclude this work by on these and listing avenues for future work.}

\begin{figure*}[bt!]
	\centering
        \includegraphics[width=7in,angle=0]{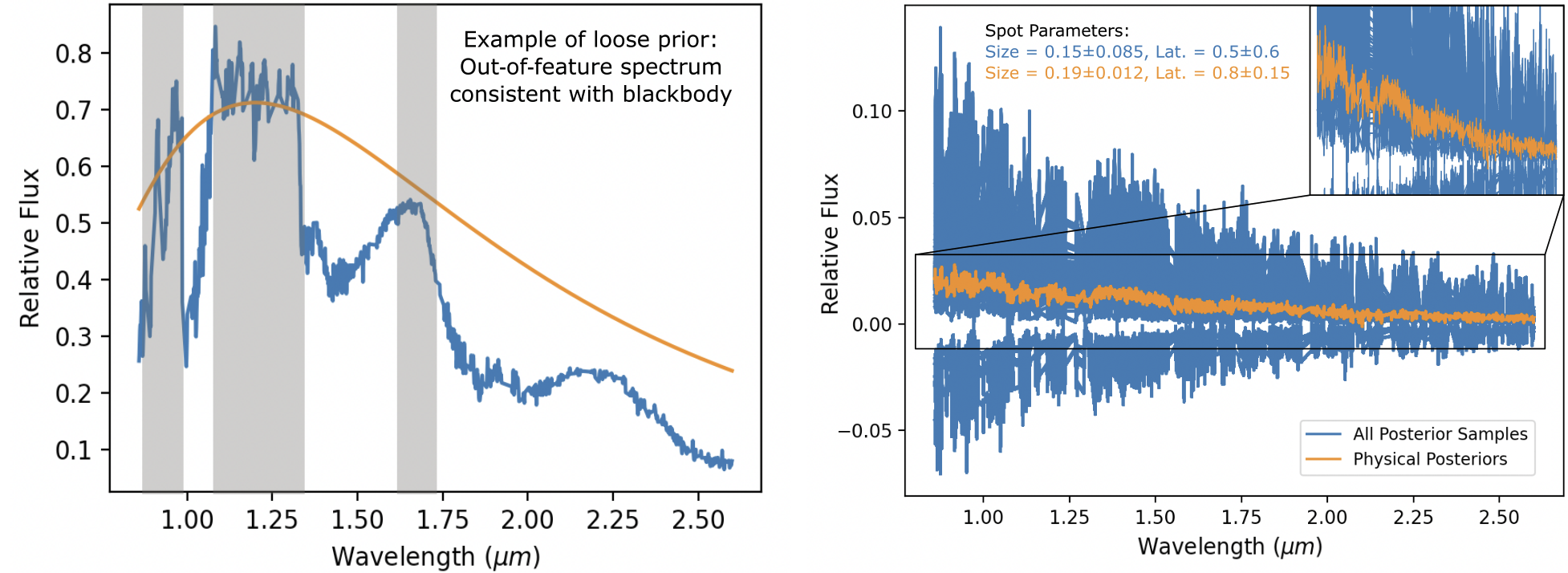}
	\caption{\R{On the use of loose (non-biasing) physical constraints. (Left) Example of constraining the out-of-feature portion of a spectrum (grey bands) to be consistent with a blackbody envelope. 
 (Right)  All spectra from a blind retrieval for a spot shown in blue, with the subset consistent with a blackbody envelope shown in orange. All spectra have been multiplied by the projected surface area of the spot feature to illustrate scatter due only to model uncertainty. Using this simple physics-based constraint, precision in the retrieved spectra is increased by $\sim15\times$, which leads to an increase in the precision on the spot size and position by $\sim5$--$8\times$. \RR{The true spot size and latitude for this example are 0.2 and 0.7 radians respectively, illustrating that the posterior constraints provide a more accurate and precise retrieval.} }}
	\label{fig: bb scaling}
\end{figure*}

\subsection{Correcting for Contamination at Different Epochs}

\RRR{The ultimate goal of this work is to generate a library of empirically derived spectra for heterogeneities (i.e. spots) of a given star, to be used in the robust correction of in-transit stellar contamination at any past or future epoch.}
 The feasibility of this approach is supported by the following. First, heterogeneities of a given star have been shown to have consistent properties. For example, molecular-band modeling of echelle spectra of DM UMa suggests a spot temperature of $3570 \pm 100$\,K during an observing campaign in 1995, with filling factors ranging from $0.25 \pm 0.08$ to $0.30 \pm 0.10$ \citep{o'neal:1998}. Returning to the same star during six nights in 1998, a later analysis found a spot temperature of $3450 \pm 120$\,K and filling factors ranging from $0.28 \pm 0.06$ to $0.42 \pm 0.05$ \citep{o'neal:2004}. Second, properties of heterogeneities appear to be correlated\R{,} making \R{them} easier to pin down. Starspot temperatures show a clear dependence on photospheric temperature, based on Doppler imaging, modeling of molecular bands, and atomic line-depth ratios \citep{berdyugina:2005}. Therefore while heterogeneit\R{ies'} filling factors surely evolve over a stellar activity cycle, their temperatures and thus spectra are a static characteristic of a given star\R{,} supporting our proposition of their relevance across epochs.

\subsection{Challenges \& Opportunities with Limb Darkening}

In the present work, we use limb darkening laws which were independant of wavelength and temperature. In practice however, limb darkening is an effect which depends on both the wavelength and the temperature of the stellar surface \citep{claret:2011}, which are related to quantities we are attempting to retrieve in the first place. Thus we find ourselves in a loop where the stellar spectrum is required in order to know the appropriate value of the limb darkening coefficients, which is required in order to fit for the stellar spectrum. 

\R{The assumption of temperature and wavelength independence is necessary for the present performance estimates, but does not lead to an overestimation of such performances. Indeed, by making said assumptions, we are in fact discarding information content--that when accounted for in future works--will further help constrain the stellar properties. Accounting for the wavelength dependency of limb darkening will notably further help break the  degeneracies currently observed between the latitude and size of a heterogeneity and thus better constrain the latitudinal distribution of heterogeneities \citep{juvan:2018,luger:2021}.}

\subsection{\R{On Complementary Physics-based Constraints}}

\R{In the current framework, the retrieval is performed in a manner that is completely agnostic to the underlying physics used to generate the synthetic data, and imposes no constraints that the retrieved spectra adhere to any physical processes. When performing a completely blind retrieval (i.e., imposing no constraints on the geometry of spot features), the model will allow a range of solutions, including those with negative or non-physical spectra for the spots. Future works should explore how relevant priors could be added to the framework without introducing biases from limited stellar models.}

An example of how physical constraints may be used to inform a retrieval without injecting bias from physical models is to impose the condition that the fitted spectra must be consistent with a blackbody outside of absorption features. We show this in \autoref{fig: bb scaling}, where we have taken the ensemble of posterior spectra of a surface feature, scaled them by the projected area of the feature, and discarded those which are not well fit by a black-body envelope in given wavelength ranges. We do not fit for a specific temperature; rather, \RRR{we discard posteriors which have a reduced 
$\chi^2$ value above 5 to allow for fits that deviate due to spectral features which may be realistic but unaccounted for by a blackbody profile, while rejecting i.e. spectra with negative values etc.}. The result of this is the reduction in uncertainty on the spot spectrum by a factor of $\sim 15\times$, as well as reducing the uncertainty on the spot size by a factor of $\sim7\times$ and the latitude by a factor of $\sim4\times$. We highlight that since this cut is made on posterior spectra, we are not injecting bias into the retrieval of the spectra themselves, rather discarding those results which do not agree with a loose physical constraint.

Further considerations to include physical information would be a parametrization of the relative flux expected between wavelength bins associated to the feature of a same molecule. While absolute flux values may be biased, relationships between wavelengths may be robust enough to provide additional constraints. This information could be extracted using Gaussian processes in order to measure correlations between different wavelengths \citep{perger:2023}. Constraining the spectra in this way would enable tighter constraints on the size and latitude of a given feature, which is currently degenerate with the overall amplitude of its spectrum. Additionally, including the use of activity indicators provided by high-precision spectroscopy to help solve in the inverse problem of reconstructing active regions on the stellar surface \citep{mallonn:2018,rosich:2020}.

In other words, while a series of improvements to this framework can (and should) be made in the future, the present theoretical proof-of-concept suffices to move towards a practical application with JWST data as a next step. Such data would also inform in a relevant manner the aforementioned series of improvements (e.g., empirical wavelength- and temperature-dependencies of the limb-darkening). We thus look forward to an on-sky validation and further development of this framework in the near future to enable the robust atmospheric characterization of planets whose spectra would otherwise stay contaminated.

\section{Acknowledgements}
We thank Elsa Ducrot and the Pandora Team for helpful discussions regarding this project.
B.V.R. thanks the Heising-Simons Foundation for support.
This material is based upon work supported by the National Aeronautics and Space Administration under Agreement No.\ 80NSSC21K0593 for the program ``Alien Earths''.
The results reported herein benefited from collaborations and/or information exchange within NASA’s Nexus for Exoplanet System Science (NExSS) research coordination network sponsored by NASA’s Science Mission Directorate.

\bibliography{ms}

\appendix{}

\restartappendixnumbering

\section{Test Model Sample Parameters \& Results}

\begin{figure*}[t!]

\begin{center}
\includegraphics[width=6in,angle=0]{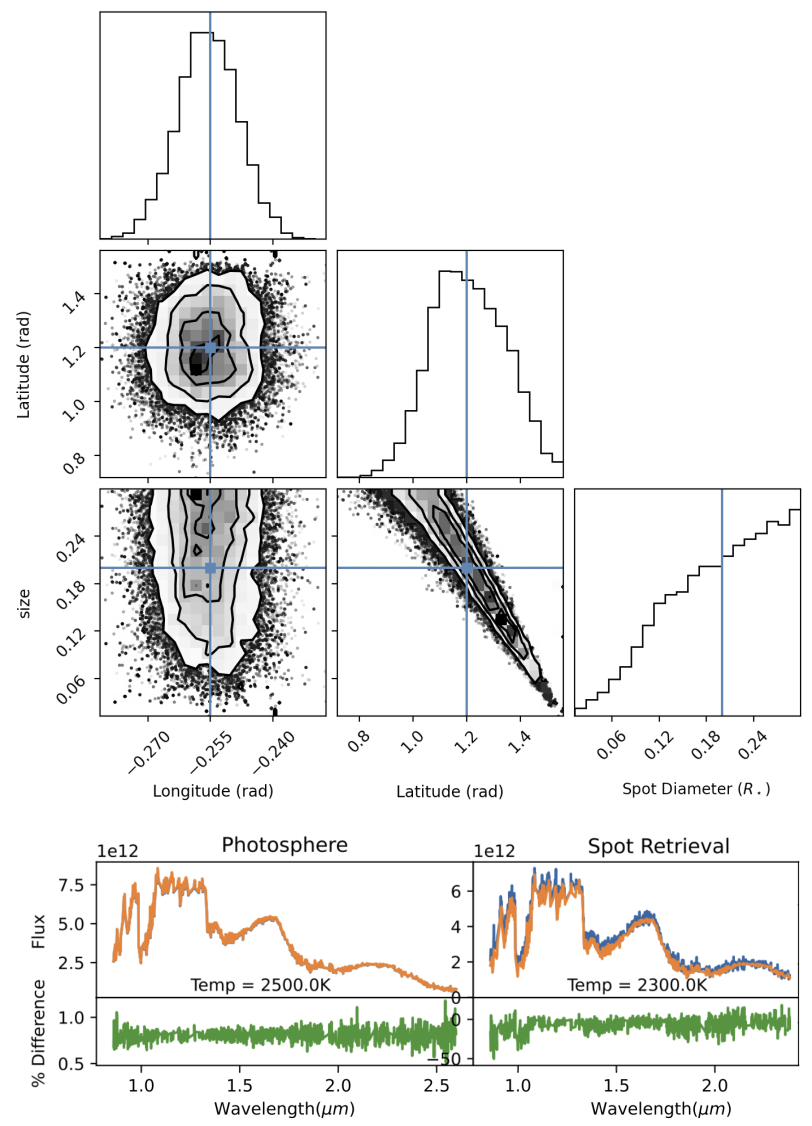}
	\caption{\RR{An example retrieval when fitting for the position, size, and spectral contribution of a spot feature. The upper plot shows the posterior distribution for the position and size of the spot feature, with the true value indicated in blue. The bottom panel shows the retrieved spectrum of both the photosphere and the spot itself, similar to \autoref{fig:spectra_fits}.}}
	\label{fig: real fit}
\end{center}
\end{figure*}

\RR{
\autoref{tab: model features} gives the parameters of the different stellar surfaces used to test the retrieval framework.
\autoref{fig: real fit} illustrates the results of the test retrieval allowing the position, size, and spectral contribution of a spot feature to all vary.
}

\begin{table*}[ht!]
\vspace{3pt}
\resizebox{\textwidth}{!}{
\hskip-4.0cm\
\begin{tabular}{c c c c |c c c c |c c c c |c c c c || c c c c c }
$\theta_1$ & $\phi_1$ & $r_1$ & $T_1$ & 
$\theta_2$ & $\phi_2$ & $r_2$ & $T_2$ & 
$\theta_3$ & $\phi_1$ & $r_3$ & $T_3$ &
$\theta_4$ & $\phi_4$ & $r_4$ & $T_4$ &
$\sigma_p$ & $\sigma_1$ & $\sigma_2$ & $\sigma_3$ & $\sigma_4$ \\
\hline
2.26 & -1.34 & 0.31 & 2700.0 & 1.01 & -1.01 & 0.11 & 2700.0 & 1.27 & -0.26 & 0.09 & 2300.0 & -1.68 & 0.66 & 0.2 & 2700.0 & 0.19 & 11.89 & 80.08 & 106.87 & 4.6\\
-1.25 & -0.01 & 0.08 & 2700.0 & -1.14 & -0.56 & 0.24 & 2700.0 & 1.58 & -0.57 & 0.12 & 2700.0 & 0.95 & 0.3 & 0.14 & 2700.0 & 0.06 & 49.94 & 8.23 & 7.49 & 4.09\\
-2.32 & 1.12 & 0.13 & 2300.0 & 2.84 & 1.28 & 0.09 & 2700.0 & 3.04 & 0.32 & 0.15 & 2300.0 & -2.36 & 0.66 & 0.08 & 2700.0 & 0.05 & 357.09 & 167.68 & 31.97 & 254.66\\
2.65 & 0.91 & 0.09 & 2700.0 & 0.48 & -1.33 & 0.19 & 2700.0 & -1.02 & -0.45 & 0.22 & 2700.0 & 1.19 & 1.12 & 0.11 & 2300.0 & 0.11 & 23.92 & 16.43 & 2.12 & 40.85\\
-1.51 & -0.57 & 0.34 & 2300.0 & 3.11 & 1.43 & 0.11 & 2700.0 & 0.07 & -0.06 & 0.06 & 2300.0 & 2.06 & -0.25 & 0.2 & 2300.0 & 0.24 & 2.45 & 118.47 & 82.94 & 5.77\\
\hline
-1.72 & -1.36 & 0.08 & 2700.0 & 0.19 & -1.03 & 0.12 & 2700.0 & 2.97 & -1.22 & 0.2 & 2300.0 & & & & & 0.11 & 70.83 & 14.42 & 16.5& \\
-0.31 & 0.89 & 0.3 & 2700.0 & -0.98 & -0.45 & 0.24 & 2300.0 & 1.16 & -1.23 & 0.12 & 2300.0 & & & & & 0.05 & 1.59 & 3.23 & 35.74& \\
0.6 & -0.58 & 0.27 & 2300.0 & 1.47 & 0.59 & 0.11 & 2700.0 & -1.41 & 0.03 & 0.22 & 2700.0 & & & & & 0.06 & 1.8 & 6.94 & 1.29& \\
-2.71 & -0.25 & 0.2 & 2700.0 & 0.72 & -1.19 & 0.17 & 2700.0 & 1.34 & 1.49 & 0.11 & 2700.0 & & & & & 0.07 & 1.87 & 11.7 & 140.47& \\
-1.87 & -0.86 & 0.24 & 2700.0 & -1.14 & 1.11 & 0.23 & 2700.0 & -1.47 & -0.13 & 0.27 & 2300.0 & & & & & 0.05 & 6.63 & 14.95 & 11.11& \\
\hline
-1.0 & 1.36 & 0.16 & 2300.0 & -1.91 & 0.55 & 0.13 & 2700.0 & & & & & & & & & 0.04 & 31.72 & 4.23& & \\
2.75 & 0.82 & 0.08 & 2300.0 & -1.43 & -0.41 & 0.18 & 2700.0 & & & & & & & & & 0.05 & 28.56 & 2.05& & \\
-1.62 & 1.19 & 0.3 & 2300.0 & -2.32 & -0.95 & 0.32 & 2700.0 & & & & & & & & & 0.04 & 5.69 & 1.37& & \\
-2.78 & -0.36 & 0.34 & 2300.0 & -0.2 & 1.48 & 0.25 & 2700.0 & & & & & & & & & 0.11 & 1.24 & 22.2& & \\
0.79 & 1.19 & 0.24 & 2300.0 & -0.6 & 0.21 & 0.09 & 2700.0 & & & & & & & & & 0.04 & 5.91 & 5.98& & \\
\hline
0.0 & 0.0 & 0.1 & 2700.0 & & & & & & & & & & & & & 0.04 & 4.43& & & \\
0.0 & 0.39 & 0.1 & 2700.0 & & & & & & & & & & & & & 0.04 & 4.92& & & \\
0.0 & 0.79 & 0.1 & 2700.0 & & & & & & & & & & & & & 0.04 & 7.93& & & \\
0.0 & 1.18 & 0.1 & 2700.0 & & & & & & & & & & & & & 0.04 & 20.51& & & \\
\hline
0.0 & 0.0 & 0.2 & 2700.0 & & & & & & & & & & & & & 0.04 & 1.14& & & \\
0.0 & 0.39 & 0.2 & 2700.0 & & & & & & & & & & & & & 0.04 & 1.48& & & \\
0.0 & 0.79 & 0.2 & 2700.0 & & & & & & & & & & & & & 0.04 & 1.92& & & \\
0.0 & 1.18 & 0.2 & 2700.0 & & & & & & & & & & & & & 0.04 & 4.83& & & \\
\hline
0.0 & 0.0 & 0.05 & 2700.0 & & & & & & & & & & & & & 0.04 & 17.85& & & \\
0.0 & 0.0 & 0.11 & 2700.0 & & & & & & & & & & & & & 0.04 & 4.55& & & \\
0.0 & 0.0 & 0.17 & 2700.0 & & & & & & & & & & & & & 0.04 & 1.62& & & \\
0.0 & 0.0 & 0.24 & 2700.0 & & & & & & & & & & & & & 0.04 & 0.9& & & \\
\hline
0.0 & 0.79 & 0.05 & 2700.0 & & & & & & & & & & & & & 0.04 & 33.53& & & \\
0.0 & 0.79 & 0.11 & 2700.0 & & & & & & & & & & & & & 0.04 & 6.35& & & \\
0.0 & 0.79 & 0.17 & 2700.0 & & & & & & & & & & & & & 0.04 & 2.72& & & \\
0.0 & 0.79 & 0.24 & 2700.0 & & & & & & & & & & & & & 0.04 & 1.47& & & \\
\hline
\end{tabular}
}
\caption{Parameters for the different stellar surfaces used to test the retrieval framework where $\theta$ and $\phi$ are the latitude and longitude of a spot feature in radians, $r$ is its size (relative to the stellar radius), and $T$ is the temperature of the model (in Kelvin) used to generate the synthetic observations. The $\sigma$ parameters represent the median fractional deviation of the measured spectra from input spectra. Solid horizontal lines separate families of models. The first three are a random sampling of models with 4, 3 and 2 spot features, with maximal spot sizes of $r = 0.3$ and temperatures of either 2300K or 2700K. The fourth and fifth group show models with a single spot of varying latitudes, for two fixed sizes. The last two groups are for a series of models with spots of varying sizes, at two fixed latitudes. }
\label{tab: model features}
\end{table*}

\end{document}